\documentclass[11pt]{article}
\usepackage{epsfig}
\usepackage{dcolumn}
\usepackage{bm}
\usepackage{graphicx}
\usepackage[textsize=tiny]{todonotes}
\usepackage{subcaption}
\usepackage{here}
\usepackage{amssymb,amsmath}
\usepackage{multirow}
\usepackage{cite,color,url}
\usepackage{tabu}
\usepackage{array}
\usepackage[colorlinks=true,urlcolor=blue,anchorcolor=blue
,citecolor=blue,filecolor=blue,linkcolor=blue,menucolor=blue
,linktocpage=true,pdfproducer=medialab,pdfa=true]{hyperref}

\usepackage{colordvi}
\usepackage{epsfig,psfrag,rotating,soul}
\def\beqa{\begin{eqnarray}}
\def\eeqa{\end{eqnarray}}

\evensidemargin \oddsidemargin
\allowdisplaybreaks

\psfrag{lk}{$l_k$}
\psfrag{lm}{$l_m$}
\psfrag{blk}{$\bar {l}_k$}
\psfrag{blm}{$\bar {l}_m$}
\def\thefootnote{\fnsymbol{footnote}}

\let\OLDthebibliography\thebibliography
\renewcommand\thebibliography[1]{
\OLDthebibliography{#1}
\setlength{\parskip}{0pt}
\setlength{\itemsep}{0pt plus 0.3ex}}
\usepackage{makecell}

\begin{document}

\thispagestyle{empty}
\begin{center}
\begin{Large}
\textbf{\textsc{Single charged Higgs boson production at the LHC}}
\end{Large}

\vspace{1cm}
{
A. Arhrib$^{3}$%
\footnote{\tt \href{mailto:aarhrib@gmail.com}{aarhrib@gmail.com}}
R. Benbrik,$^{2}$%
\footnote{\tt
\href{mailto:r.benbrik@uca.ma}{r.benbrik@uca.ma}}
M. Krab,$^{1}$%
\footnote{\tt
\href{mailto:mohamed.krab@usms.ac.ma}{mohamed.krab@usms.ac.ma}}	
B. Manaut,$^{1}$%
\footnote{\tt
\href{mailto:b.manaut@usms.ma}{b.manaut@usms.ma}}
M. Ouchemhou,$^{2}$%
\footnote{\tt
\href{mailto:mohamed.ouchemhou@ced.uca.ac.ma}{mohamed.ouchemhou@ced.uca.ac.ma}}
} and
Qi-Shu Yan$^{4,5}$%
\footnote{\tt
\href{mailto:yanqishu@ucas.ac.cn}{yanqishu@ucas.ac.cn}}

\vspace*{.7cm}

{\sl
$^1$ Research Laboratory in Physics and Engineering Sciences, Modern and Applied Physics
Team, Polidisciplinary Faculty, Beni Mellal, 23000, Morocco}\\
{\sl
$^2$ Polydisciplinary Faculty, Laboratory of Fundamental and Applied Physics,
Cadi Ayyad University, Sidi Bouzid, B.P. 4162, Safi, Morocco}\\
{\sl
$^3$ Abdelmalek Essaadi University, Faculty of Sciences and techniques,
Tanger, Morocco}\\
{\sl
$^{4}$ Center for Future High Energy Physics, Chinese Academy of Sciences, Beijing 100049, PR
China.}\\
{\sl
$^{5}$ School of Physics Sciences, University of Chinese Academy of Sciences, Beijing 100039, PR
China.}
\end{center}

\vspace*{0.1cm}

\begin{abstract}
A search for charged Higgs may yield clear and direct signs of new physics outside the realm of the Standard Model (SM). In the Two-Higgs Doublet Model (2HDM), we investigate two of the main single charged Higgs production channels at the Large Hadron Collider (LHC), assuming that either $h$ or $H$ replicates the detected resonance at $\sim 125~\rm{GeV}$. We consider the possibility of the charged Higgs boson production through the $pp \rightarrow H^\pm W^\mp$ and $pp \rightarrow  H^\pm bj$ production processes that may present additional significance at the LHC experiments. Considering the $H^\pm \rightarrow W^\pm h_i/A$ decay channels and mainly concentrating on the $b\bar{b}$, $\tau\tau$ and $\gamma\gamma$ decays of $h_i$ and $A$, we examine the possible signatures arising from the aforementioned charged Higgs production and decay in both type-I and type-X realizations of 2HDM. Our study shows that these signatures can have sizable rates at low $\tan\beta$ as long as the condition $M_{H^\pm} < m_t - m_b$ is satisfied.
As a result, we propose the $bb$, $\tau\tau$ and $\gamma\gamma$ final states associated with $WW$ or $Wbj$ as an encouraging experimental avenue that would complement the LHC search for a charged Higgs boson.
\end{abstract}
 
\def\thefootnote{\arabic{footnote}}
\setcounter{page}{0}
\setcounter{footnote}{0}
\section{Introduction}
\label{intro}
The discovery of the Higgs boson was made possible through the CERN Large Hadron Collider (LHC). The first observation of such a particle was reported by the two collaborations ATLAS and CMS \cite{ATLAS:2012yve,CMS:2012qbp} in 2012. Regarding particle content, this is the ultimate and one of the most significant discoveries inside the Standard Model (SM) of particle physics.
Although, due to many theoretical and experimental reasons, the SM Higgs boson is widely assumed not to be the only fundamental particle with spin zero. Many theories extending the SM exist and predict the existence of a new charged particle(s) with spin zero. Direct searches for this charged Higgs boson(s) have already been conducted at LEP \cite{ALEPH:2013htx}, Tevatron \cite{CDF:2011pxh}, and now at the LHC \cite{ATLAS:2021upq,CMS:2020imj}, all yielding negative results. As a consequence, depending on the charged Higgs boson mass, constraints are placed on
the branching ratios of different decay channels for the charged Higgs boson. 
The LEP experiment excludes the mass of charged Higgs below $80~\rm{GeV}$ at 95$\%$ CL, considering only the decays $H^+ \to c\bar{s}$ and $H^+ \to \tau \nu_{\tau}$ with BR$(H^+ \to c\bar{s}) + \mathrm{BR}(H^+ \to \tau \nu_{\tau}) = 1$ \cite{LEPHiggsWorkingGroupforHiggsbosonsearches:2001ogs}. Such a bound gets stronger if BR$(H^+ \to \tau \nu_{\tau}) = 1$ \cite{Logan:2009uf}. 
For a charged Higgs mass of $100$ GeV, searches at the Tevatron based on top anti-top pair production with the subsequent $t \to bH^+$ decay (setting BR$(H^+ \to \tau^+ \nu) = 1$) have set a limit on BR$(t \to bH^+)$ to be less than 0.2 \cite{D0:2009hbc}. 
LHC searches for $H^\pm$ have set an upper limit at $95\%$ confidence level on the production cross section multiplied by the branching ratio, $\sigma(pp \rightarrow H^\pm tb)\times \rm{BR}(H^\pm \rightarrow tb)$, which ranges from 3.6 (2.6) pb at $M_{H^\pm} = 200$ GeV to $0.036~(0.019)$ pb at $M_{H^\pm} = 2$ TeV. The discovery prospects of the charged Higgs boson in other independent frameworks can be found in Refs. \cite{Coleppa:2019cul,Coleppa:2021wjx}.

The Two-Higgs-Doublet Model (2HDM) is a simplest beyond SM framework that predicts the charged Higgs bosons, where an additional complex doublet $\Phi_2$ was added to the SM Higgs sector. In order to prevent the Flavor Changing Neutral Currents (FCNCs) at the tree-level \cite{Diaz-Cruz:2020pjf}, a $Z_2$ symmetry was introduced $(\Phi_1 \to \Phi_1, \Phi_2 \to -\Phi_2)$, leading to four distinct interaction modes \cite{Barger:1989fj}, when it is expanded to include the model's fermions, known as in this context, type-I, type-II, type-X (or lepton-specific) and, type-Y (or flipped). These models are widely discussed in the literature, and both direct and indirect constraints are set on them. The former is already set above and excludes the charged Higgs boson with a mass less than $80$ GeV at $95\%$ CL, while the latter is model-dependent and dominated by the B-physics decay, mainly through, $B \rightarrow X_s\gamma$ \cite{Haller:2018nnx}. A light charged Higgs boson with a mass below $100$ GeV is still allowed in type-I and type-X as long as $\tan\beta$ is larger than 2. However, such constraints are very strong in type-II and type-Y, excluding a charged Higgs mass below $680$ GeV. The last update on this constraint $B \rightarrow X_s \gamma$ \cite{Misiak:2020vlo} excludes $M_{H^\pm}$ below $800$ GeV in type-II and type-Y, while type-I and type-X are still accommodated the light charged Higgs Boson below $100$ GeV.

At the LHC, the hunt for the charged Higgs boson were performed with many distinct production modes, starting with the most promising one, which is the $t\bar{t}$ production and decay, which represents an excellent source of $H^\pm$ when $M_{H^\pm} < m_t-m_b$. The top (anti-top) quark could decay into $H^+ b~(H^- \bar{b})$, competing with the SM decay of $t \rightarrow W^+ b~(\bar{t}\to W^- \bar{b})$. The production process $pp \rightarrow t\bar{t} \rightarrow b\bar{b} H^- W^+ + \rm{C.C.}$ has a sizable cross section that can serve as a significant supply of light charged Higgs bosons. In addition, the following production modes might be used to look for light charged Higgs at the LHC: associated production with top and bottom quarks considering either the $gg \rightarrow t\bar{b} H^+$ process in the four-flavor scheme or the $g\bar{b} \rightarrow tH^+$ process in the five-flavor scheme \cite{Alwall:2004xw,Bahl:2021str}, associated production with a $W$ boson through $b\bar{b} \rightarrow H^\pm W^\mp$, which is dominated at tree-level, and $gg \rightarrow H^\pm W^\mp$ that is dominated at loop-level \cite{Moretti:1998xq,BarrientosBendezu:1998gd,BarrientosBendezu:1999vd}, associated production with a bottom quark and a light quark \cite{Arhrib:2015gra,Moretti:1996ra}, resonant production via the quark anti-quark collision $c\bar{s},c\bar{b} \rightarrow H^+$ \cite{Hernandez-Sanchez:2012vxa,Hernandez-Sanchez:2020vax,Dittmaier:2007uw}, associated production with a neutral Higgs states $h_i$ or $A$, $q\bar{q}' \rightarrow H^{\pm} h_i/A$ \cite{Kanemura:2001hz, Akeroyd:2003bt, Akeroyd:2003jp, Cao:2003tr, Belyaev:2006rf, Miao:2010rg}, where $h_i$ stand for $h$ and $H$, as well as the pair production through the annihilation process $q\bar{q} \rightarrow H^\pm H^\mp$ or gluon fusion $gg \rightarrow H^\pm H^\mp$ \cite{BarrientosBendezu:1999gp, Brein:1999sy, Moretti:2001pp,Moretti:2003px}.

This work aims to examine the production of single charged Higgs boson along
with a $W$ boson as well as with a bottom quark and a light jet, i.e. $pp \rightarrow H^\pm W^\mp$ and $pp \rightarrow H^\pm bj$, in the 2HDM type-I and type-X frameworks. These models still predict a light charged Higgs boson with significant rates of its bosonic decays that potentially dominate over fermionic modes. We study the different possible LHC signatures stemming from the aforementioned Higgs production channels and the bosonic decays $H^\pm \rightarrow W^\pm h_i/A$ as well as their phenomenological implications in the context of the LHC.  

This paper is structured as follows. We briefly introduce the 2HDM framework in Section \ref{2hdm_model}. In Section \ref{sect:constraints}, we outline the theoretical and experimental constraints that will be forced on 2HDM parameter space during our scan.  In Section \ref{sect:process}, we study the charged Higgs production at the LHC. Our LHC signatures are discussed in Section \ref{sect:signatures} and our conclusions are given in Section \ref{sect:conclusion}.
\section{The two-Higgs doublet model}
\label{2hdm_model}
Upon the soft broken $Z_2$ symmetry, the scalar potential that is $SU(2)_L\otimes U(1)_Y$ invariant and CP-conserving is given by:
\begin{align}
V_{\rm{2HDM}}(\Phi_1,\Phi_2) ~=~& m_{11}^2(\Phi_1^+\Phi_1)+m_{22}^2(\Phi_2^+\Phi_2)-m_{12}^2(\Phi_1^+\Phi_2+ \rm{h.c.})\nonumber\\
&+\lambda_1(\Phi_1^+\Phi_1)^2+\lambda_2(\Phi_2^+\Phi_2)^2+\lambda_3(\Phi_1^+\Phi_1)(\Phi_2^+\Phi_2)\nonumber\\
&+\lambda_4 (\Phi_1^+\Phi_2)(\Phi_2^+\Phi_1)+\frac{\lambda_5}{2}[(\Phi_1^+\Phi_2)^2+ \rm{h.c.}], \label{RTHDMpot}
\end{align}
where all parameters are real valued due to the CP-conserving requirement. $m_{11}^2$, $m_{22}^2$ and $m_{12}^2$ are mass parameters, and $\lambda_{1-5}$ are coupling parameters without dimensions. Spontaneous Electro-Weak Symmetry Breaking (EWSB) allows $\Phi_1$ and $\Phi_2$ doublets to gain vacuum expectation values denote as $v_{1,2}$, which satisfy $v_1^2 + v_2^2 = v^2 \approx (246 ~\rm{GeV})^2$. In addition, during the (EWSB), we left with ten free independent parameters: $m_{11}^2,~m_{22}^2,~m_{12}^2,~v_1,~v_2$ and $\lambda_{1-5}$. 
Through the two minimization conditions of the potential, replacing $m_{11}^2$ and $m_{22}^2$ by $v_{1,2}$, we are left with the seven free independent, real, parameters: 
$\lambda_{1-5},~m_{12}^2,~\tan\beta(=v_2/v_1)$. Instead, for more convenience, we adopt the following physical parameters:
$M_h,~M_H,~M_A,~M_{H^\pm},~\alpha,~\tan\beta,~m_{12}^2$. 
The angle $\beta~(\alpha)$ is the rotation angle from the group eigenstate to the mass eigenstate in the charged Higgs and the CP-odd (CP-even) Higgs sector.

In order to suppress FCNCs, the Yukawa sector of general 2HDM is constrained by $Z_2$ symmetry, which implies that each fermion type is only allowed to interact with one of the Higgs doublets $\Phi_{1,2}$. Such a requirement gives rise to four different versions of Yukawa interactions carrying the names 2HDM type-I, type-II, type-X and, type-Y. In type-I, the mass of fermions is generated by the doublet $\Phi_2$.
In type-II, leptons and down quarks receive mass from the doublet $\Phi_1$, while up quarks receive mass from the doublet $\Phi_2$. 
In type-X, the quarks interact with $\Phi_2$, while the charged leptons interact with $\Phi_1$.
In type-Y, down quarks interact with $\Phi_1$, whereas the rest of the fermion types (up-type quarks and leptons) interact with $\Phi_2$.
Here, we only target the 2HDM type-I and type-X.

In the mass eigenstate basis, the interactions between the fermion and Higgs sector, is described by the Yukawa Lagrangian \cite{Branco:2011iw},
\begin{align}
- {\mathcal{L}}_{\rm Yukawa} = \sum_{f=u,d,l} \left(\frac{m_f}{v} \xi_f^h \bar{f} f h + 
\frac{m_f}{v}\xi_f^H \bar{f} f H 
- i \frac{m_f}{v} \xi_f^A \bar{f} \gamma_5 f A \right) + \nonumber \\
\left(\frac{V_{ud}}{\sqrt{2} v} \bar{u} (m_u \xi_u^A P_L +
m_d \xi_d^A P_R) d H^+ + \frac{ m_l \xi_l^A}{\sqrt{2} v} \bar{\nu}_L l_R H^+ + \rm{h.c.} \right),
\label{Yukawa-1}
\end{align}
where $V_{ud}$ represents a CKM matrix element. the coefficients $\xi_f^{h_i, A}$ are 2HDM Higgs couplings to fermions normalized to the SM couplings, which are listed in Table \ref{coupling}.
		\begin{table}[H]
			\begin{center} 
				\begin{tabular}{|c||c|c|c|c|c|c|c|c|c|} \hline
					&$\xi_{u}^{h}$&$\xi_{d}^{h}$&$\xi_{l}^{h}$&$\xi_{u}^{H}$&$\xi_{d}^{H}$&$\xi_{l}^{H}$&$\xi_{u}^{A}$&$\xi_{d}^{A}$&$\xi_{l}^{A}$\\\hline
                    type-I & $c_\alpha/s_\beta$ & $c_\alpha/s_\beta$& $c_\alpha/s_\beta$ & $s_\alpha/s_\beta$ & $s_\alpha/s_\beta$ & $s_\alpha/s_\beta$ & $c_\beta/s_\beta$ & 
                    $-c_\beta/s_\beta$ & $-c_\beta/s_\beta$ \\ \hline
                    type-X & $c_\alpha/s_\beta$ & $c_\alpha/s_\beta$& $-s_\alpha/c_\beta$ & $s_\alpha/s_\beta$ & $s_\alpha/s_\beta$ & $c_\alpha/c_\beta$ & $c_\beta/s_\beta$ & 
                    $-c_\beta/s_\beta$ & $s_\beta/c_\beta$ \\ \hline
				\end{tabular}
			\end{center}
			\caption{\label{coupling}Yukawa couplings of 2HDM Higgs bosons to the fermions.}
		\end{table}

\section{Parameter space scans and constraints}
\label{sect:constraints}
We randomly scan the parameters of the 2HDM type-I and type-X using the public code \texttt{2HDMC-1.8.0} \cite{Eriksson:2009ws}, considering both Normal Scenario (NS) and Inverted Scenario (IS), i.e. the observed $125$ GeV Higgs boson at the LHC is assigned to either $\mathcal{CP}$-even states $h$ (NS) or $H$ (IS), in the ranges illustrated in Table \ref{param_scans}.
We require each point to be subjected to the updated experimental and theoretical constraints.

\begin{table}[H]
	\centering
	\setlength{\tabcolsep}{4.3pt}
	\begin{tabular}{|c||c|c|c|c|c|c|c|}\hline
		 &$M_h~[\mathrm{GeV}]$&$M_H~[\mathrm{GeV}]$&$M_A~[\mathrm{GeV}]$&	$M_{H^\pm}~[\mathrm{GeV}]$& $\sin(\beta-\alpha)$&$\tan\beta$&$m_{12}^2~[\mathrm{GeV}^2] $ \\\hline
		NS	&$125.09$&$[126;\,700]$&$[15;\,700]$&$[80;\,700]$& $[0.95;\,1]$&$[2;\,25]$&$[0;\,m_H^2\cos\beta\sin\beta]$\\\hline
		IS	&$[15;\,120]$&$125.09$&$[15;\,700]$&$[80;\,700]$& $[-0.5;\,0.5]$&$[2;\,25]$&$[0;\,m_h^2\cos\beta\sin\beta]$\\\hline
	\end{tabular}		
	\caption{2HDM type-I and type-X input parameters.} \label{param_scans}
\end{table}

From the theoretical side, the perturbativity, vacuum stability, and unitarity constraints are enforced as the following: 
\begin{itemize}
\item Vacuum stability to get a global minimum and not only a local vacum. We should force the following conditions \cite{Barroso:2013awa,Deshpande:1977rw},
\begin{align}
\lambda_1 > 0,\quad\lambda_2>0, \quad\lambda_3>-\sqrt{\lambda_1\lambda_2} ,\quad \lambda_3+\lambda_4-|\lambda_5|>-\sqrt{\lambda_1\lambda_2}.
\end{align}

\item Unitarity constraints require that the scattering amplitudes of scalar-scalar(vector) and vector-vector to be unitary at high energies. This requirement reflects certain limits on the eigenvalues of the S-matrix of such scattering processes as follows $e_i\leq 4\pi$~\cite{Kanemura:1993hm,Akeroyd:2000wc,Arhrib:2000is}, which in turn, affect the potential parameter by the requirement $e_i = f(\lambda_i)$.

\item Perturbativity constraints demand to all quartic scalar couplings to be perturbative, i.e. to fulfill the limit $\lambda_i\leq 4\pi$\cite{Chang:2015goa}.

\end{itemize}
The parameters that survive the theoretical constraints above are furthermore asked to satisfy the following experimental constraints:
\begin{itemize}
\item The electroweak oblique parameters $S$ and $T$ (setting $U=0$) \cite{Grimus:2007if,Grimus:2008nb} are imposed to set limits
on the mass separation between the 2HDM Higgs states using the following fit result \cite{ParticleDataGroup:2020ssz}:
\begin{align}
&S=0.05 \pm 0.08,\quad T=0.09 \pm 0.07,\quad \rho_{ST}=0.92,
\end{align}
where $\rho_{ST}$ is the correlation coefficient between $S$ and $T$. 
\item Constraints from additional Higgs bosons searches at collider experiments are imposed using the tool \texttt{HiggsBounds-5.10.2} \cite{Bechtle:2020pkv}.
\item Agreement with SM-like Higgs signal measurements are tested using \texttt{HiggsSignals-2.6.2} \cite{Bechtle:2020uwn}. We enforce to the predicted Higgs signal strengths to be compatible with the experimental measurements at $95\%$ CL. 

\item Flavor physics constraints are checked using the tool \texttt{SuperIso v4.1} \cite{Mahmoudi:2008tp}. Relevant observables are listed bellow \cite{Haller:2018nnx}.
\begin{itemize}
\item BR($B\to X_{s}\gamma)_{E_\gamma\ge 1.6\ \text{GeV}}$ = $(3.32\pm 0.15)\times10^{-4}$,
\item BR($B_{s}\to \mu^{+}\mu^{-}$) = $(3.0 \pm 0.6 \pm 0.25) \times 10^{-9}$,
\item BR($B_{\mu}\to \tau\nu$) = $(1.06 \pm 0.19) \times 10^{-4}$.
\end{itemize}

\end{itemize}

In order to fulfill the discussed constraints, especially those from the parameters $S$ and $T$, $M_H$ , $M_A$ and $M_{H^\pm}$ are required to be close to each other. Such a requirement is sensitive to the $T$ parameter which strongly correlates these masses. 
In the NS, either $M_A \sim M_{H^\pm}$ or $M_H \sim M_{H^\pm}$ must be satisfied leading to an alternative dominance of the decays $H^\pm \rightarrow W^\pm A$ or $H^\pm \rightarrow W^\pm H$ (competing $H^\pm \rightarrow tb$ decay) in large parts of the parameter space. The decay $H^\pm \rightarrow W^\pm A$ dominates until $M_A \sim 125$ GeV, then a strong competition occurs above this threshold between $tb$ and $WH$ decays.
In the IS, $M_{H^\pm}$ and $M_A$ cannot be too large. In type-I, we find $M_{H^\pm} \lesssim 643$ GeV and $M_A \lesssim 659$ GeV, while in type-X $M_{H^\pm} \lesssim 637$ GeV and $M_A \lesssim 634$ GeV. (We should note that these upper limits are not absolute. These limits might be exceeded by further, allowed, points.) In this scenario, the dominated decays are $H^\pm \rightarrow W^\pm h$ or $H^\pm \rightarrow W^\pm A$ depending on $M_A$, which plays a crucial role to separate these decays. The decay channel $H^\pm \rightarrow W^\pm A$ dominates only at low $M_A$, i.e. when $M_A \lesssim 100~\rm{GeV}$. Once we cross this threshold, $Wh$ decay becomes the dominant one since it enjoys more phase space.

\section{Charged Higgs production}
\label{sect:process}
\subsection{$H^\pm W^\mp$ production}
\label{sect:results}
\begin{figure}[H]
	\centering
	\begin{tabular}{cccc}
		\begin{subfigure}[t]{.2\textwidth}\centering
			\includegraphics[height=1.5cm,width=2.5cm]{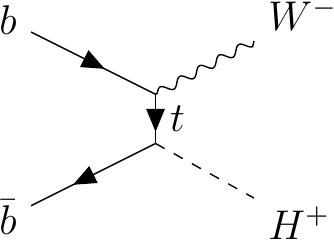}
			\caption{}
			\label{fig:Hptb_prod_bb_nonres}
		\end{subfigure}
		\hfill
		&\begin{subfigure}[t]{.2\textwidth}\centering
			\includegraphics[height=1.5cm,width=2.5cm]{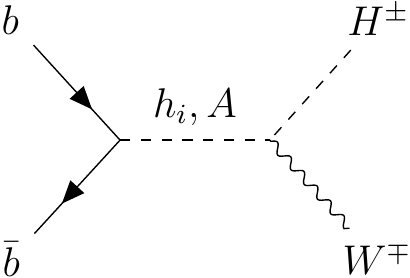}
			\caption{}
			\label{fig:Hptb_prod_bb_res}
		\end{subfigure}
		
		&\begin{subfigure}[t]{.2\textwidth}\centering
			\includegraphics[height=1.5cm,width=2.5cm]{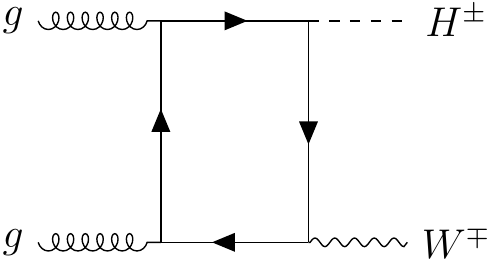}
			\caption{}
			\label{fig:Hptb_prod_gg_nonres}
		\end{subfigure}
		\hfill
		&\begin{subfigure}[t]{.2\textwidth}\centering
			\includegraphics[height=1.5cm,width=2.5cm]{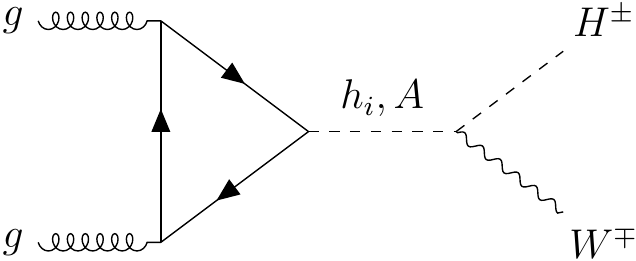}
			\caption{}
			\label{fig:Hptb_prod_gg_res}
		\end{subfigure}
	\end{tabular}
	\caption{Feynman diagrams contributing to the $H^\pm W^\mp$ production.}
	\label{fig:Hptb_prod_feynman}
\end{figure}

One of the important charged Higgs production channels, which merits particular attention, is its associated production with a $W$ boson.
The Feynman diagrams for such a channel are modeled in four separate subprocesses, as illustrated in Fig. \ref{fig:Hptb_prod_feynman}, each contributing to the $H^\pm W^\mp$ production cross section.
The $b\bar{b}$ contribution comes with two subprocess, the non-resonant (Fig. \ref{fig:Hptb_prod_bb_nonres}) and resonant (Fig. \ref{fig:Hptb_prod_bb_res}) channels, while the two remaining contributions are given by the $gg$ initiated non-resonant and resonant channels, which are depicted in Figs. \ref{fig:Hptb_prod_gg_nonres} and  \ref{fig:Hptb_prod_gg_res}, respectively.

Many studies are devoted to the $pp \rightarrow H^\pm W^\mp$ production process in literature. The calculation of the leading order cross section can be found in Refs. \cite{Dicus:1989vf,BarrientosBendezu:1999vd,BarrientosBendezu:1998gd,Brein:2000cv,Bao:2010sz}. Radiative corrections beyond leading order have been calculated in Refs.\cite{Hollik:2001hy,Dao:2010nu,Enberg:2011ae,Kidonakis:2017dmh}. 
Phenomenological studies can be found in Refs.  \cite{Moretti:1998xq,Asakawa:2005nx,Eriksson:2006yt,Hashemi:2010ce,Bao:2011sy,Aoki:2011wd, Bahl:2021str}.  The experimental prospect of seeing $W^\pm H^\mp$ synthesis at the LHC with the subsequent $H^- \to \bar{t}b$ decay\cite{Moretti:1998xq} and $H^- \to \tau^- \bar{v}_{\tau}$ decay\cite{Eriksson:2006yt,Hashemi:2010ce} has also been studied. 


At the tree-level, the dominant production channel is the $b\bar{b}$ contribution, while the $gg$ contribution dominates at one loop-level. The cross sections for these contributions\footnote{We neglect the $b\bar{b}$ resonant channel (Fig. \ref{fig:Hptb_prod_bb_res}) since the Yukawa couplings are small in both types of interest. The resonant $gg$ contribution is only relevant when $M_{h_{\rm{BSM}}} > M_{H^\pm} + M_W$ or $M_{A} > M_{H^\pm} + M_W$. While the constraints from oblique parameters favors the mass degeneracy, especially for heavy $H^\pm$, only one of the aforementioned conditions can take place. And since we focus on the $H^\pm \rightarrow W^\pm h_{\rm{BSM}}/A$ decays, the $gg$ resonant channel (Fig. \ref{fig:Hptb_prod_gg_res}) is no longer relevant in our scenario. Such a channel, $gg \rightarrow A \rightarrow H^\pm W^\mp$, strongly competes with $gg \rightarrow A \rightarrow Z h$ channel near the alignment limit.} are calculated at the leading order using the code \texttt{MadGraph@aMC$\_$NLO} \cite{Alwall:2014hca}, considering $\sqrt{s}=14$ TeV and using the pdf set MMHT2014 \cite{Harland-Lang:2014zoa}.
Note that the approximate NNLO (aNNLO) correction enhances the cross section $\sigma{(b\bar{b} \rightarrow H^\pm W^\mp})$ by a factor of 1.38 \cite{Kidonakis:2017dmh,Bahl:2021str}.

\begin{figure}[H]
	\centering
	\includegraphics[width=0.4\textwidth]{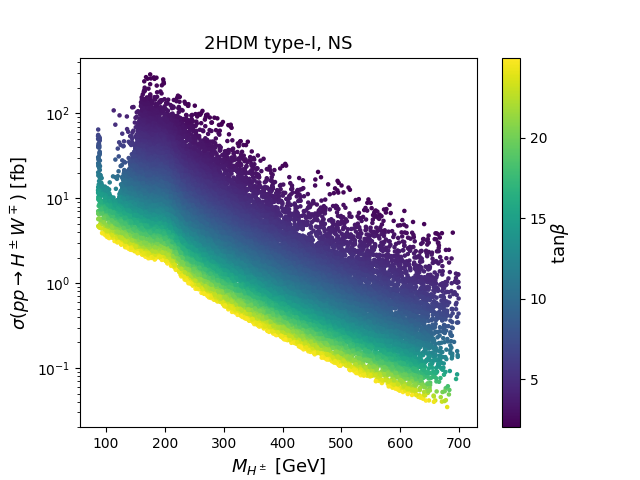}
	\includegraphics[width=0.4\textwidth]{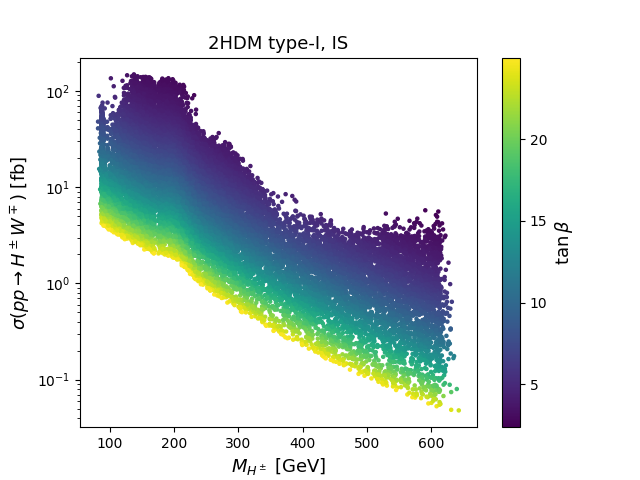} \\
	\caption{Cross section for $pp \to H^\pm W^\mp$ production process, at $\sqrt{s}=14$ TeV, as a function of $M_{H^\pm}$, with $\tan\beta$ indicated by the color map.}\label{fig-xsppHcW-type1}
\end{figure}

The scatter points in Fig. \ref{fig-xsppHcW-type1} illustrates the $pp \rightarrow H^\pm W^\mp$ production cross section as a function of $M_{H^\pm}$, in NS (left panels) and IS (right panels), with the color map shows $\tan\beta$.
One can see in both NS and IS that these cross sections reach their maximum at low values of $\tan\beta$. This is explained by the $tbH^\pm$ coupling, which is proportional to $1/\tan\beta$. The cross section is, indeed, independent of the considered scenario. The only difference comes from the theoretical and experimental constraints on each scenario. This production process has almost the same behavior in both NS and IS, which dominates at $H^\pm$ mass of approximately below $200~\rm{GeV}$. In the NS, $\sigma(pp \rightarrow H^\pm W^\mp)$ can reach $288$ fb, while cross section can only reach $148$ fb in the IS. These results are in type-I.
In type-X, the behavior is almost the same as in type-I. (LHC restrictions on Higgs physics are the main difference.) 
In the IS, for instance, signal cross section reaches its maximum value ($28$ fb) at around $M_{H^\pm} \sim 154$ GeV.
Note that we combined $b\bar{b}$ and $gg$ initiated channels.

\subsection{$H^\pm bj$ production}
\begin{figure}[H]
	\centering
	\begin{tabular}{cccc}
		\begin{subfigure}[t]{.2\textwidth}\centering
			\includegraphics[height=1.6cm,width=2.6cm]{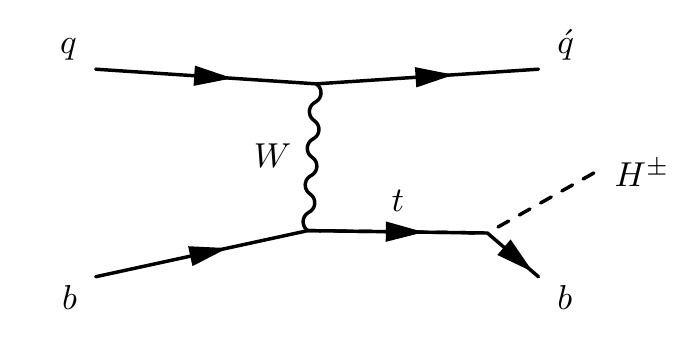}
			\caption{}
			\label{fig:Hpm_bj1}
		\end{subfigure}
		\hfill
		&\begin{subfigure}[t]{.2\textwidth}\centering
			\includegraphics[height=1.6cm,width=2.6cm]{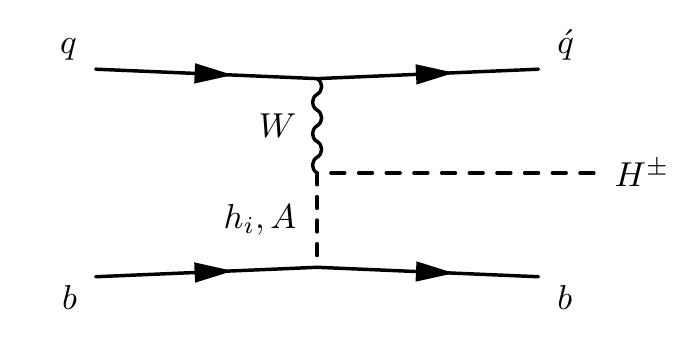}
			\caption{}
			\label{fig:Hpm_bj2}
		\end{subfigure}
		
		&\begin{subfigure}[t]{.2\textwidth}\centering
			\includegraphics[height=1.6cm,width=2.6cm]{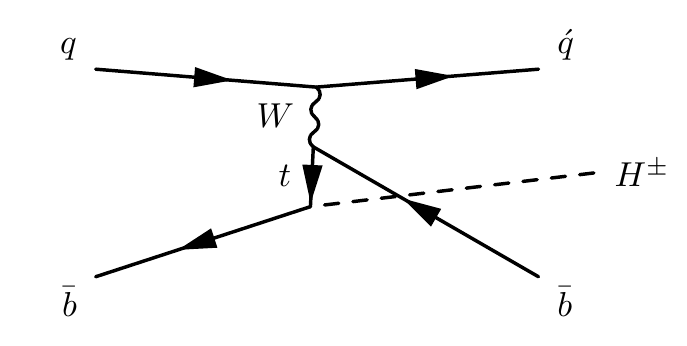}
			\caption{}
			\label{fig:Hpm_bj3}
		\end{subfigure}
		\hfill
		&\begin{subfigure}[t]{.2\textwidth}\centering
			\includegraphics[height=1.6cm,width=2.6cm]{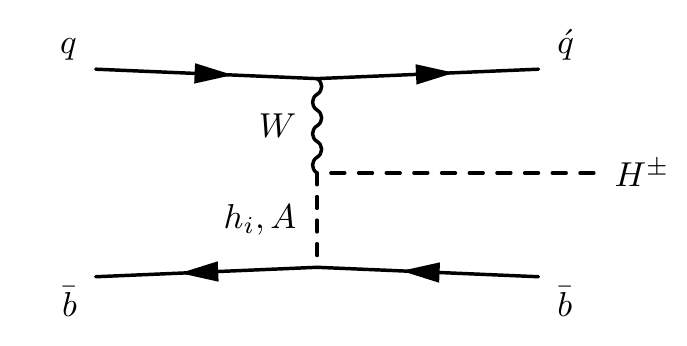}
			\caption{}
			\label{fig:Hpm_bj4}
		\end{subfigure}
	\end{tabular}
	\caption{Feynman diagrams contributing to the $H^\pm b j$ production.}
	\label{fig:Hpmbj_prod_feynman}
\end{figure}

Besides the $H^\pm W^\mp$ production discussed above, the production of
a charged Higgs boson in association with a bottom quark and a jet, i.e. $pp \rightarrow H^\pm b j$, is also interesting at small values of $\tan\beta$ in type-I and type-X of 2HDM. The Feynman diagrams for this production process are illustrated in Fig. \ref{fig:Hpmbj_prod_feynman}. 
There two main sub-processes, $b$-initiated channel ($qb \rightarrow q' H^\pm b$) and $\bar{b}$-initiated channel ($q\bar{b} \rightarrow q' H^\pm \bar{b}$). Fig. \ref{fig:Hpm_bj1} shows top s-channel exchange while Fig. \ref{fig:Hpm_bj3} illustrates $u$-channel one. Figs. \ref{fig:Hpm_bj2} and \ref{fig:Hpm_bj4} represent the $t$-channel diagrams with neutral Higgs bosons contributions.  

In literature, $pp \rightarrow H^\pm b j$ production process has been firstly studied in the context of Minimal Supersymmetric Standard Model (MSSM) \cite{Moretti:1996ra}. In this study, both $H^\pm \rightarrow \tau\nu_\tau$ and $H^\pm \rightarrow tb$ decay channels are considered, depending on the $H^\pm$ mass range.
In 2HDM, the $pp \rightarrow H^\pm b j$ production rate and its sensitivity to $\tan\beta$ has been discussed in Ref. \cite{Arhrib:2015gra}. Within the same framework, a parton level analysis (assuming single top production with the subsequent $t \rightarrow H^+ b$ decay) can be found in Ref. \cite{Aoki:2011wd}. Furthermore, an analysis including parton shower and detector effects is performed in Ref. \cite{Arhrib:2020tqk}.

\begin{figure}[H]
	\centering
	\includegraphics[width=0.4\textwidth]{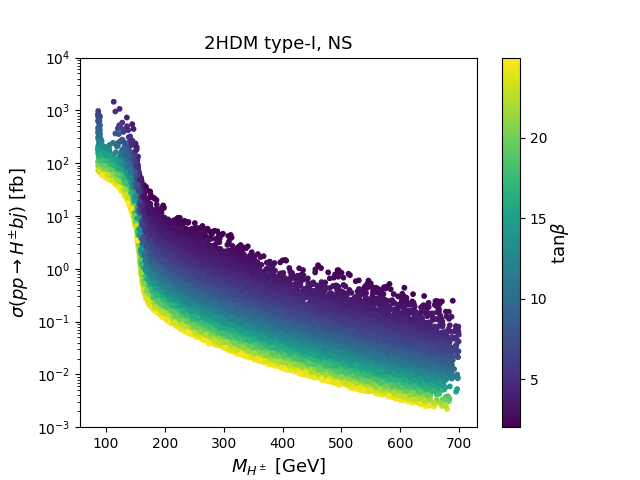}
	\includegraphics[width=0.4\textwidth]{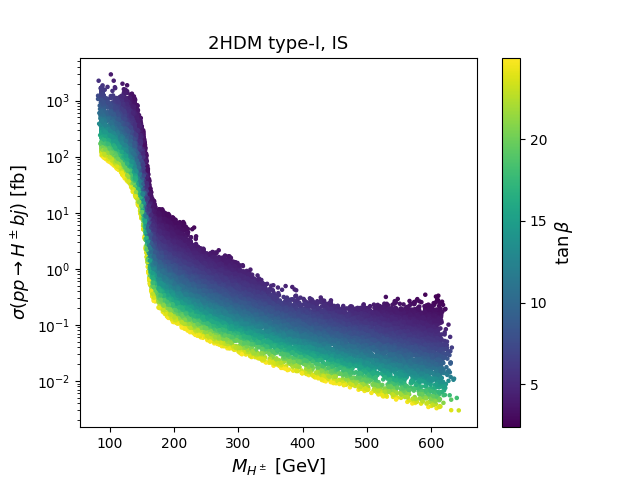}
	\caption{Cross section for $pp \to H^\pm b j$ production process, at $\sqrt{s}=14$ TeV, as a function of $M_{H^\pm}$, with $\tan\beta$ indicated by the color map.}\label{fig-xsqbHc}
\end{figure}

In Fig. \ref{fig-xsqbHc}, we show the $pp \rightarrow H^\pm b j$ production cross section as a function of $M_{H^\pm}$ in both NS (left) and IS (right), indicating $\tan\beta$ by the color map. It is visible in the plot that this cross section is increased at both small values of $\tan\beta$ and $M_{H^\pm}$, especially when top quark is produced on-shell, i.e. when $M_{H^\pm} < m_t - m_b$. In this mass region, $\sigma(pp \rightarrow H^\pm b j)$, is entirely dominated by the top s-channel contribution (Fig. \ref{fig:Hpm_bj1}), of up to $2.9$ pb can be reached in the IS, while cross section can reach values above $1$ pb in the NS. However, other diagrams (Fig \ref{fig:Hpm_bj2}) also contribute\footnote{See Ref. \cite{Arhrib:2015gra} for more discussion of each contribution.} significantly in the region where top quark is off-shell ($M_{H^\pm} > m_t - m_b$), i.e. diagrams involving $H^\pm W^\mp h_i$ and $H^\pm W^\mp A$ couplings. In the NS (IS), only diagrams including $H^\pm W^\mp H$ ($H^\pm W^\mp h$) and $H^\pm W^\mp A$ vertex are important.
It is, moreover, clear from Fig. \ref{fig-xsqbHc} that the cross section rapidly drops when $M_{H^\pm} > m_t - m_b$. This is explained by the cancellation between the top diagram and diagrams involving $H^\pm W^\mp h_i/A$ couplings. In type-X, the behavior is similar to type-I.
In the NS (IS), for instance, signal cross section reaches $142~(204)$ fb of its maximum possible value at around $M_{H\pm} \simeq 96~(99)$ GeV. 
\section{LHC signatures}
\label{sect:signatures}
In large parts of the allowed parameter space (especially in type-I of 2HDM), the charged Higgs boson decay into a $W$ boson and a neutral Higgs ($h$, $H$ or $A$) is remarkably dominated and is favored by the alignment limit. 
For the neutral Higgs boson ($h$, $H$ or $A$) decays, we mainly consider the $bb$, $\tau\tau$ and $\gamma\gamma$ channels.
In this regard, we concentrate on the following signatures:
\begin{eqnarray}
\sigma^{S}(pp \rightarrow xWW) &=& \sigma(pp \rightarrow H^\pm W^\mp \rightarrow W^\pm S W^\mp \rightarrow x W^\pm W^\mp), \\
\sigma^{S}(pp \rightarrow x Wbj) &=& \sigma(pp \rightarrow H^\pm bj \rightarrow W^\pm S bj \rightarrow x W^\pm bj),
\end{eqnarray}
where $S$ can be either $h$, $H$ or $A$, and $x$ stands for $bb$, $\tau\tau$ or $\gamma\gamma$. We then expect two $W$ bosons and a pair of bottom quarks, tau leptons or photons as signatures\footnote{Following the same philosophy, the charged Higgs boson production can also provide four-bottom, two-bottom and two-tau, four-tau as well as four-photon, with one or two $W$ bosons, final states \cite{Arhrib:2017wmo,Arhrib:2020tqk,Bahl:2021str,Arhrib:2021xmc,Wang:2021pxc, Arhrib:2021yqf, Cheung:2022ndq, Arhrib:2022inj}.} of charged Higgs boson.  
\subsection*{Part-I: Normal scenario}
\begin{figure}[H]
	\centering
	\begin{tabular}{cc}
		\includegraphics[width=0.4\textwidth]{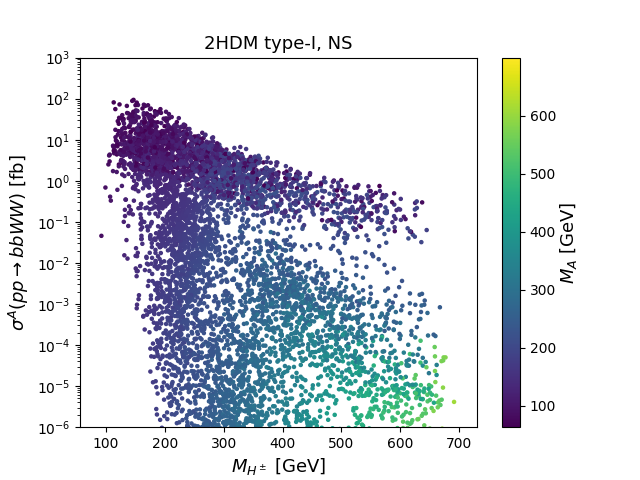} &
		\includegraphics[width=0.4\textwidth]{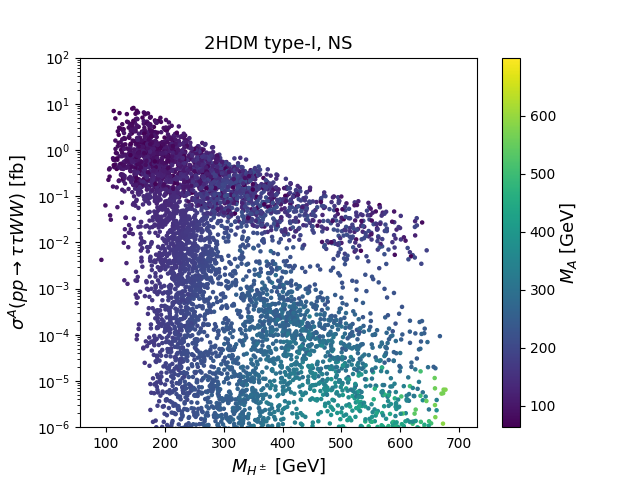} \\
		\includegraphics[width=0.4\textwidth]{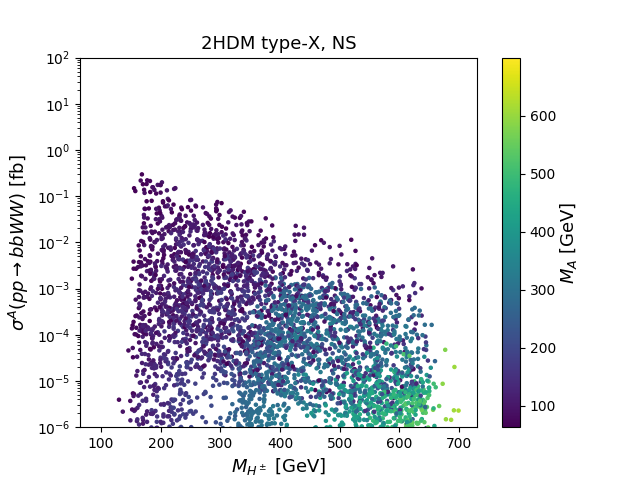} &
		\includegraphics[width=0.4\textwidth]{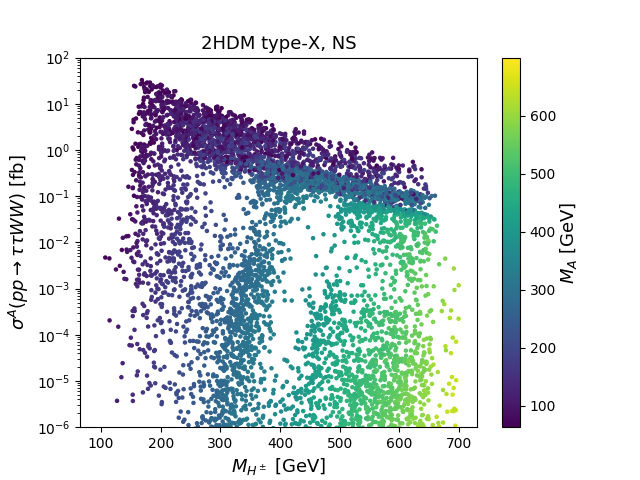} 
	\end{tabular}
	\caption{$\sigma^{A}{(pp \rightarrow bbWW)}$ (left panel), $\sigma^{A}{(pp \to \tau\tau WW)}$ (right panel) as a function of $M_{H^\pm}$, with the color code showing $M_A$. Upper (lower) panels present the type-I (type-X) results.} \label{HcWH_A:signaturetpx}
\end{figure}

In the NS, the observed 125 signals at the LHC are attributed to the light CP-even Higgs boson $h$. 
Without the need to tune particular parameters, this scenario displays a collider phenomenology that is present over most of the possible parameter space. 
\begin{figure}[H]
	\centering
	\begin{tabular}{cc}
		\includegraphics[width=0.4\textwidth]{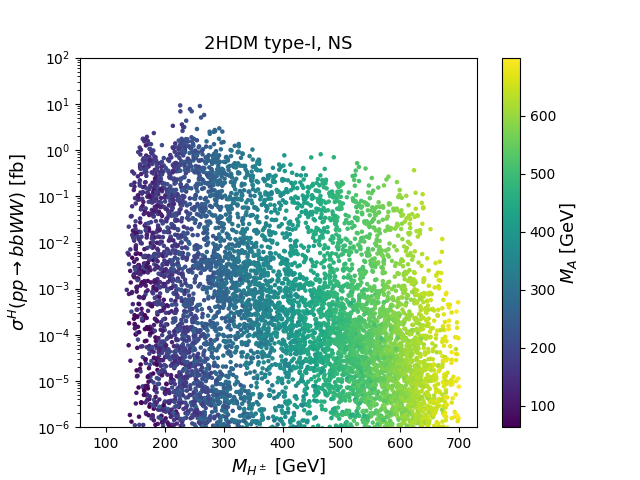} &
		\includegraphics[width=0.4\textwidth]{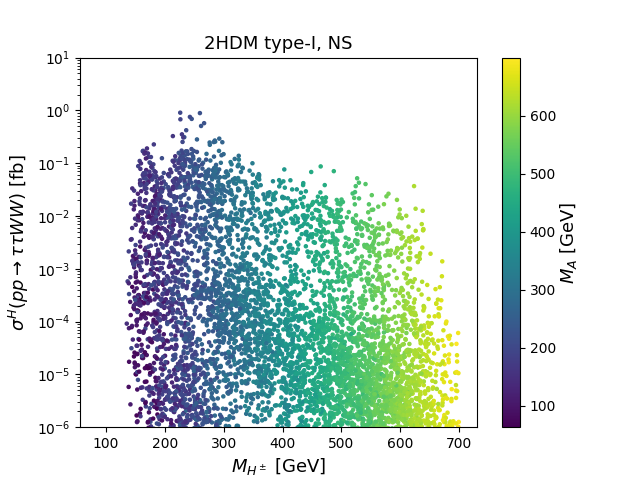} \\
		\includegraphics[width=0.4\textwidth]{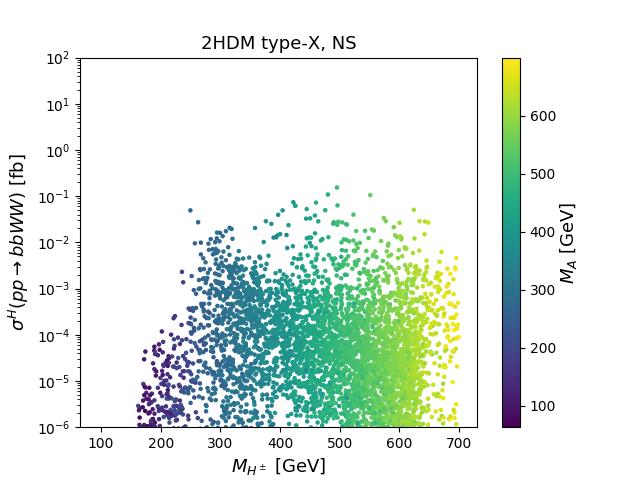} &
		\includegraphics[width=0.4\textwidth]{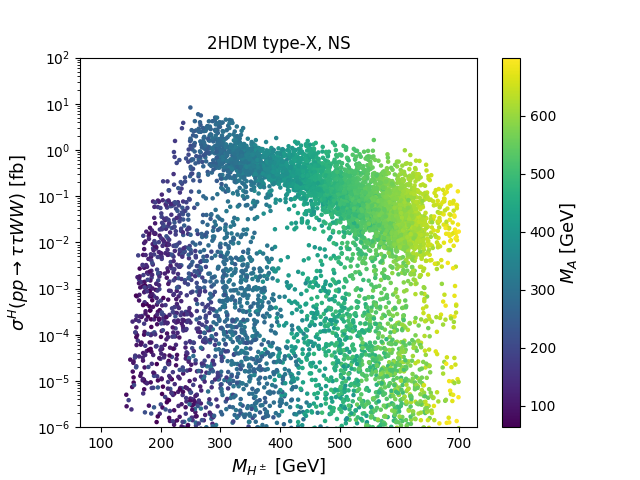} 
	\end{tabular}
	\caption{$\sigma^{H}{(pp \rightarrow bbWW)}$ (left panel), $\sigma^{H}{(pp \to \tau\tau WW)}$ (right panel) as a function of $M_{H^\pm}$, with the color code showing $M_A$. Upper (lower) panels present the type-I (type-X) results.} \label{HcWH_A:signaturetp1}
\end{figure}
In Fig. \ref{HcWH_A:signaturetp1}, we show the prediction of the final states $bbWW$ and $\tau\tau WW$ arising from $pp \rightarrow H^\pm W^\mp$ as a function of the $M_{H^\pm}$ and $M_A$ masses. In the upper panels, we depict the total signal cross sections $\sigma^H(pp \rightarrow bbWW)$ (left) and $\sigma^H(pp \rightarrow \tau\tau WW$) (right) in type-I. One can see from these figures that such cross sections reach 9.2 fb as a maximum value for $bbWW$ and less than 1 fb for $\tau\tau$ at $M_{H^\pm}\sim 226$ GeV and $M_A = 200$ GeV. This is due to the fact that the $H^\pm \to W^\pm A$ decay dominates at small $M_{H^\pm}$. In the region where $\sigma(pp \rightarrow H^\pm W^\mp)$ reaches its maximum ($M_{H^\pm} \sim 175$), the $WA$ decay still dominates, competing the $tb$ decay, over $H^\pm \to W^\pm H$ mode. Above these mass regimes, where $WH$ decay also competes with the aforementioned decays, the production cross section drops to be small and any enhancement from the decay $H^\pm \to W^\pm H$ would be insufficient to increase effectively the signal cross sections. In the lower panels, the same final states are shown in type-X framework, $bbWW$ (left) and $\tau\tau WW$ (right). From these figures, one can observe that our signal cross section $\sigma^H(pp \rightarrow bbWW)$ has a small rate in type-X framework, while cross section for $\tau\tau WW$ dominates due to the $H \rightarrow \tau\tau$ decay structure in such framework, reaching 8 fb as maximum for $M_{H^\pm} \sim M_A \sim 250$ GeV.
Instead of the $H^\pm \to W^\pm H$ decay mode investigated above, Fig. \ref{HcWH_A:signaturetpx} shows the final states $bbWW$ and $\tau\tau WW$ arising from $pp \rightarrow H^\pm W^\mp$ production with the subsequent $H^\pm \rightarrow W^\pm A$ decay in both type-I and type-X as a function of the $M_{H^\pm}$ mass. In the upper panels, we present the 2HDM type-I predictions. Such signatures are interesting for the light charged and CP-odd Higgs bosons, especially below the  $H^\pm \rightarrow W^\pm A$ threshold. Our signals $bbWW$ and $\tau\tau WW$ reach values up to $93$ fb and $8$ fb, respectively, at $M_{H^\pm} \sim 147$ GeV and $M_A \sim 67$ GeV.
The dominance of $bbWW$ over $\tau\tau WW$ comes from the decay $A \to bb$ which dominates for light $A$ in type-I of 2HDM. In the lower panels, the same signals are depicted in type-X. As expected, $\tau\tau WW$ signature is the one of interest due to enhanced $A\to \tau\tau$ decay in type-X, unlike type-I. The $bbWW$ is negligible in this configuration.

\begin{figure}[H]
	\centering
	\begin{tabular}{cc}
		\includegraphics[width=0.34\textwidth]{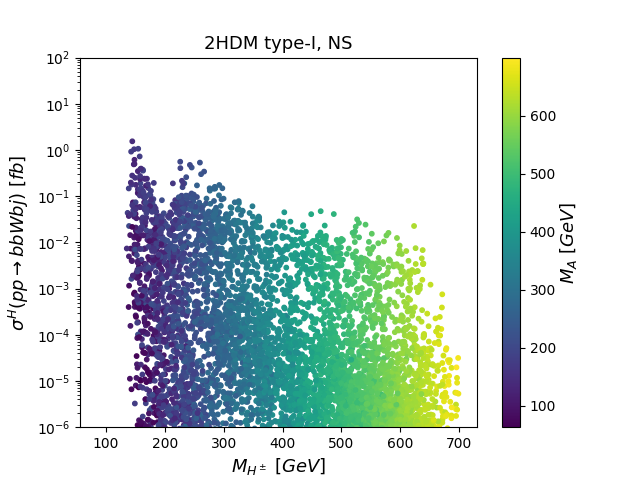} &
		\includegraphics[width=0.34\textwidth]{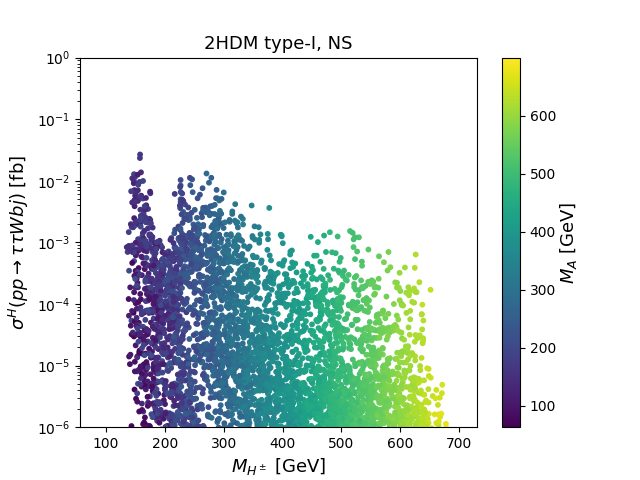}\\ 
		\includegraphics[width=0.34\textwidth]{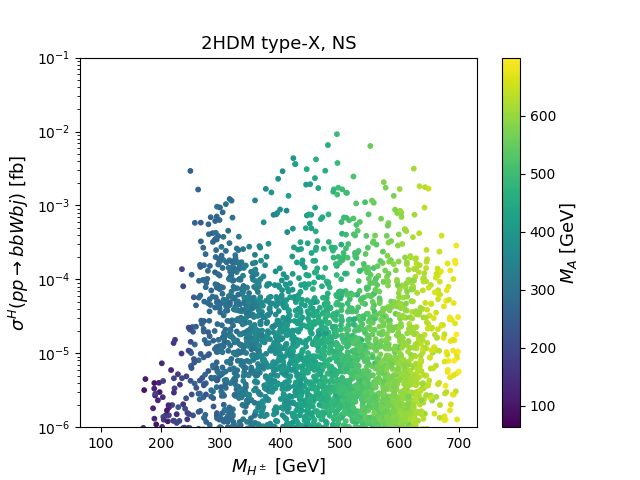} &
		\includegraphics[width=0.34\textwidth]{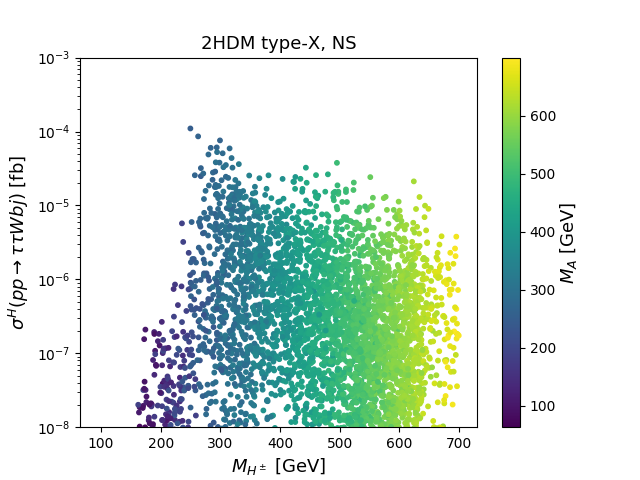} 
	\end{tabular}
	\caption{$\sigma^H{(pp \rightarrow bbWbj)}$ (left panel), $\sigma^H{(pp \to \tau\tau Wbj)}$ (right panel) as a function of $M_{H^\pm}$, with the color code showing $M_A$. Upper (lower) panels present the type-I (type-X) results.} \label{qb_hbj:signaturesH}
\end{figure}

\begin{figure}[H]
	\centering
	\begin{tabular}{cc}
		\includegraphics[width=0.34\textwidth]{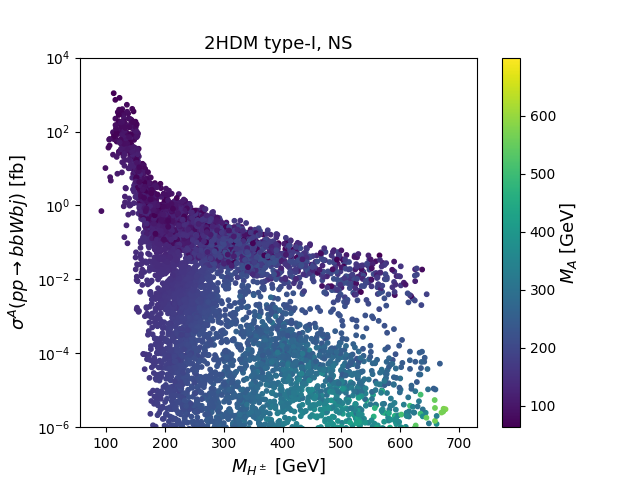} &
		\includegraphics[width=0.34\textwidth]{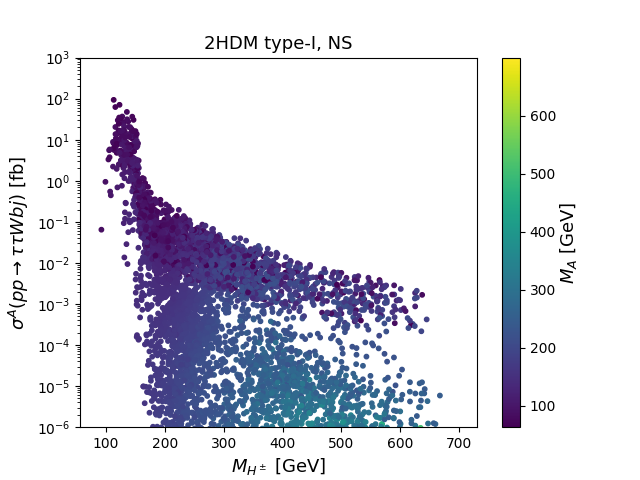}\\ 
		\includegraphics[width=0.34\textwidth]{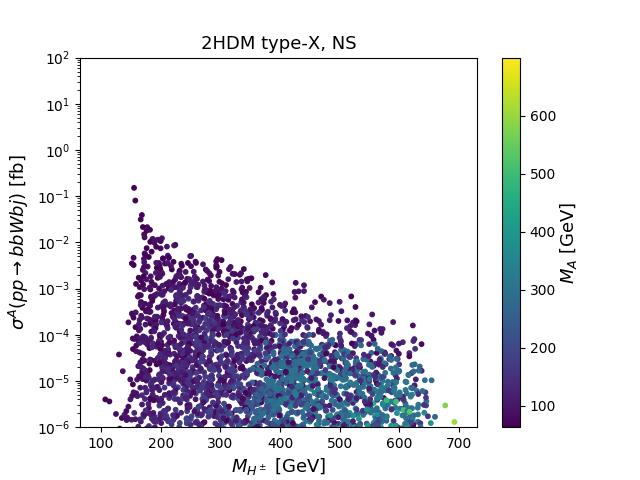} &
		\includegraphics[width=0.34\textwidth]{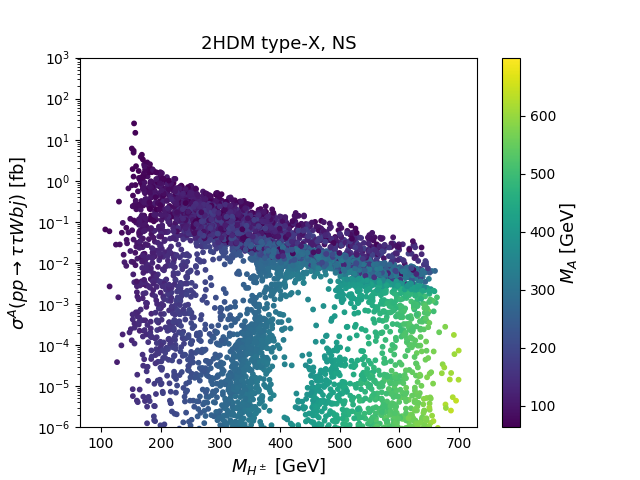} 
	\end{tabular}
	\caption{$\sigma^A{(pp \rightarrow bbWbj)}$ (left panel), $\sigma^A{(pp \to \tau\tau Wbj)}$ (right panel) as a function of $M_{H^\pm}$, with the color code showing $M_A$. Upper (lower) panels present the type-I (type-X) results.} \label{qb_hbj:signaturesA}
\end{figure}

Besides the $pp \rightarrow H^\pm W^\mp$ production process and its final states discussed above, also the $pp \rightarrow H^\pm b j$ production and its signatures can be phenomenologically interesting. Considering first the $H^\pm \rightarrow W^\pm H$ decay channel, we display in Fig. \ref{qb_hbj:signaturesH} the signal cross sections $\sigma(pp \to H^\pm b j \to W^\pm H bj \to bb W^\pm bj)$ and $\sigma(pp \to H^\pm b j \to W^\pm H bj \to \tau\tau W^\pm bj)$. These signals have small rates since the  $H^\pm \to W^\pm A$ decay channel still dominates over the $H^\pm \to W^\pm H$ channel in the area where $M_{H^\pm}<m_t-m_b$, which is the most relevant area for $H^\pm b j$ production. Above this mass region, where $WH$ decay competes with $WA$ and $tb$ decays, the cross section rates are almost comparable in these two regions even though the production cross section of $H^\pm b j$ quickly drops when $M_{H^\pm} > m_t-m_b$.
Similar to Fig. \ref{qb_hbj:signaturesH}, Fig. \ref{qb_hbj:signaturesA} depicts the cross section rates of $bbWbj$ and $\tau\tau Wbj$ considering in this case the $H^\pm \rightarrow W^\pm A$ decay instead. In type-I (upper panels), $\sigma^A(pp \rightarrow bbWbj)$ can exceed $1$ pb, while $\sigma^A(pp \rightarrow \tau\tau Wbj)$ reaches values around $94$ fb. In type-X (lower panels), the signal rate of $\tau\tau Wbj$ can reach up to $25$ fb, while the $bbWbj$ rate is negligible.
Before we turn to the inverted scenario, we point out that signatures arising from $H^\pm \rightarrow W^\pm A$ decay channel can provide the best reach compared to those from $H^\pm \rightarrow W^\pm H$ decay in the normal scenario. 

\subsection*{Part-II: Inverted scenario}
\begin{figure}[H]
	\centering
	\includegraphics[width=0.4\textwidth]{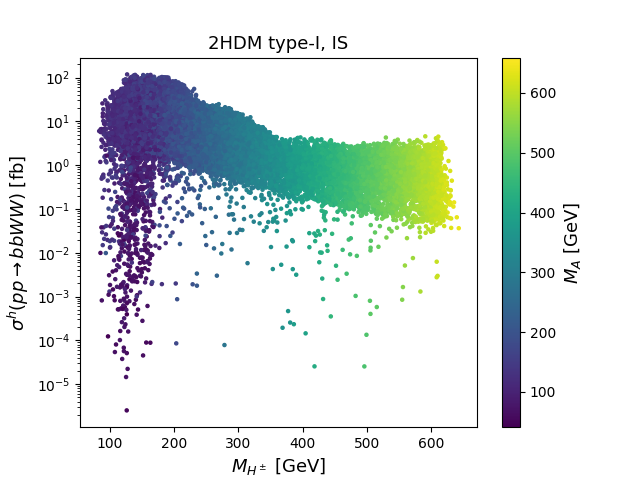} 
	\includegraphics[width=0.4\textwidth]{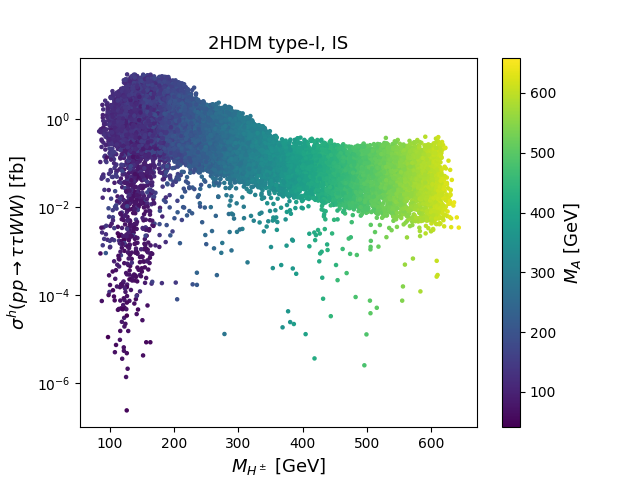} \\
	\includegraphics[width=0.4\textwidth]{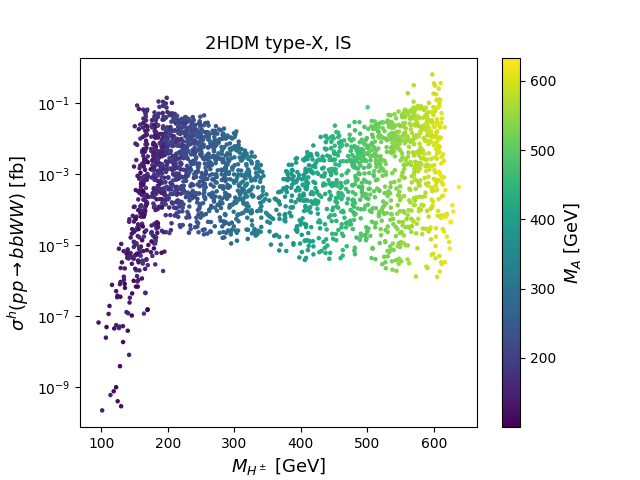} 
	\includegraphics[width=0.4\textwidth]{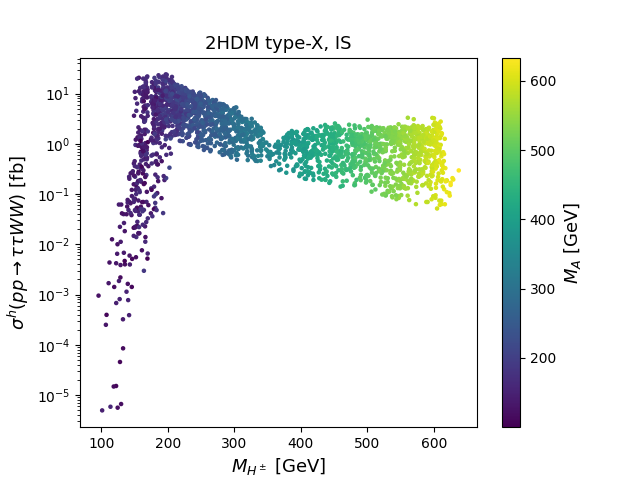} 
	\caption{$\sigma^h{(pp \rightarrow bbWW)}$ (left panel) and $\sigma^h{(pp \to \tau\tau WW)}$ (right panel) as a function of $M_{H^\pm}$, with the color code showing $M_A$. Upper (lower) panels present the type-I (type-X) results.} \label{HcW:signatures}
\end{figure}

In this scenario (IS),  we explain the 125 GeV Higgs signal by the CP-even Higgs state $H$ by fixing $M_H = 125.09$ GeV. We then sampled the remaining 2HDM parameters as illustrated in Table \ref{param_scans}. In this realization, the $H^\pm \rightarrow W^\pm h$ decay is enhanced while $H^\pm \rightarrow W^\pm H$ channel is suppressed, compared to NS. Keeping in mind this configuration, we depict in Fig. \ref{HcW:signatures} the signal cross section $\sigma^h(pp \rightarrow xWW)$ vs $M_{H^\pm}$. In the upper panels, we illustrate the $bbWW$ and $\tau\tau$ final states in type-I. These total cross sections reach their maximum only at low $A$ and $H^\pm$ masses, especially above the $H^\pm \rightarrow W^\pm A$ threshold. One sees that the cross section for $bbWW$ signature (upper-left) could reach values above $100$ fb, while the $\tau\tau WW$ signature (upper-right) reaches $10$ fb. 
In the lower panels, we present the 2HDM type-X results. One can see that our final states are interesting at intermediate and, also, large masses of $A$ and $H^\pm$. This is due to the dominance of $H^\pm \rightarrow \tau\nu_\tau$ channel at low $M_{H^\pm}$. As expected, the $\tau\tau WW$ cross section is the dominant one, reaching values above $20$ fb. The rate of $bbWW$ final state is small compared to type-I prediction. 

\begin{figure}[H]
	\centering
	\includegraphics[width=0.4\textwidth]{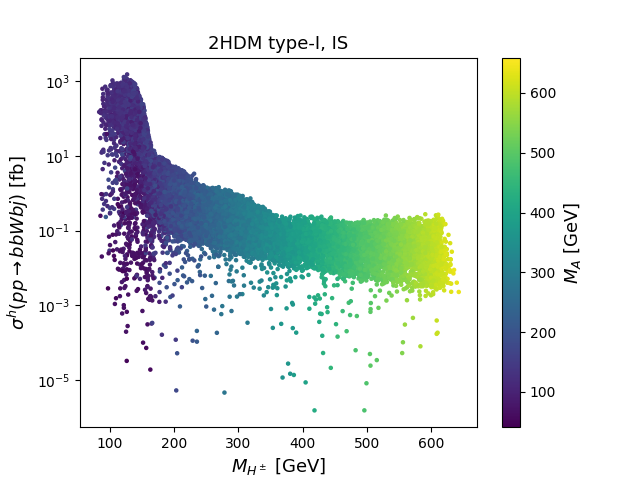} 
	\includegraphics[width=0.4\textwidth]{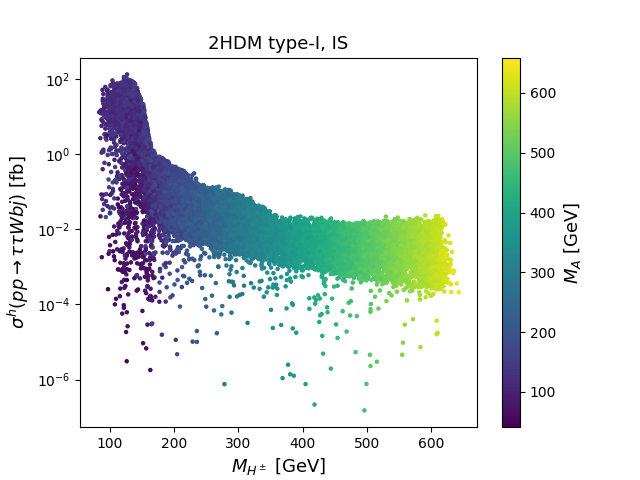} \\
	\includegraphics[width=0.4\textwidth]{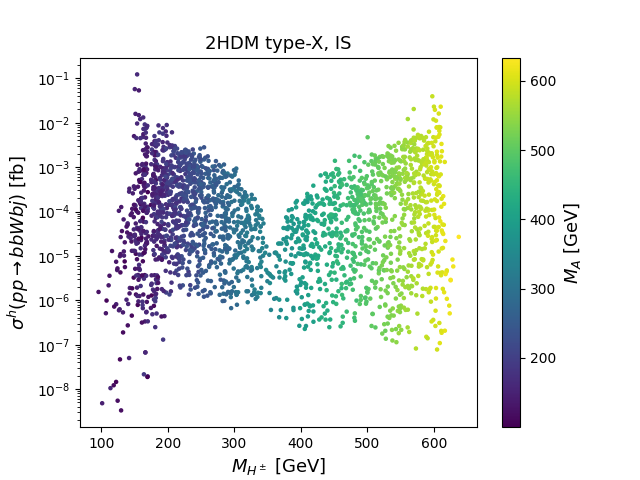} 
	\includegraphics[width=0.4\textwidth]{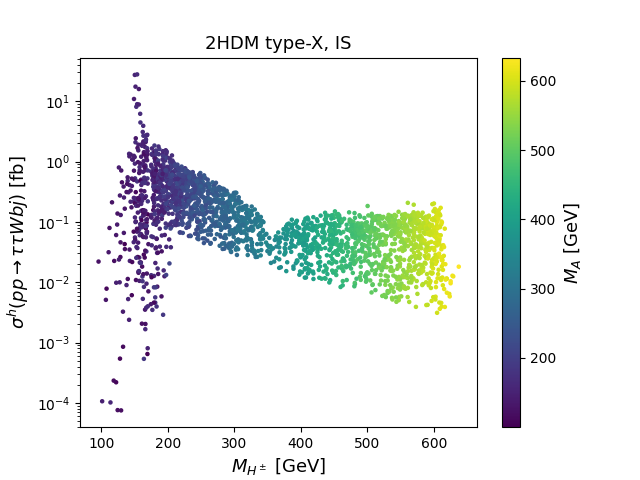} 
	\caption{$\sigma^h{(pp \rightarrow bbWbj)}$ (left panel) and $\sigma^h{(pp \to \tau\tau Wbj)}$ (right panel) as a function of $M_{H^\pm}$, with the color code showing $M_A$. Upper (lower) panels present the type-I (type-X) results.} \label{qbHc:signatures}
\end{figure}

\begin{figure}[H]
	\centering
	\includegraphics[width=0.4\textwidth]{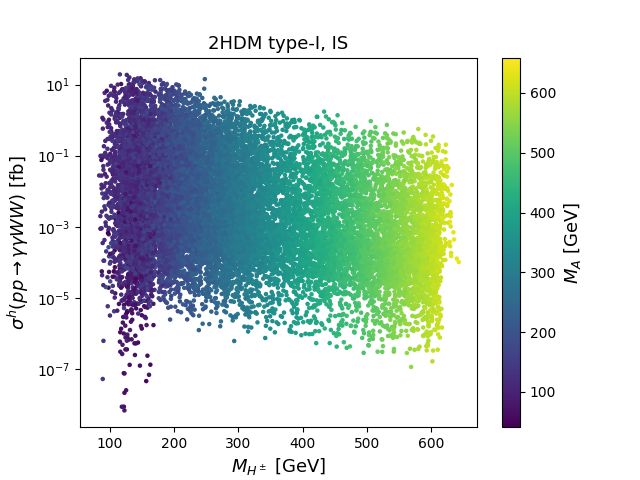} 
	\includegraphics[width=0.4\textwidth]{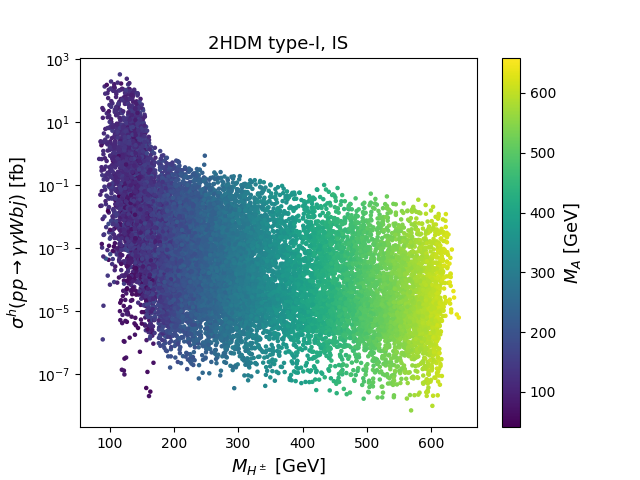} 	
	\caption{$\sigma^h{(pp \rightarrow \gamma\gamma WW)}$ (left panel) and $\sigma^h{(pp \to \gamma\gamma Wbj)}$ (right panel) as a function of the $M_{H^\pm}$ mass. The color code indicates the parameter $M_A$.} \label{gaga:signatures}
\end{figure}

In Fig. \ref{qbHc:signatures}, we present the signatures arising from $pp \rightarrow H^\pm b j$ production. The upper panels show the total cross sections $\sigma^h(pp \rightarrow bbWbj)$ (upper-left) and $\sigma^h(pp \rightarrow \tau\tau Wbj)$ (upper-right) vs $M_{H^\pm}$ in type-I framework. It is clear that $\sigma^h(pp \rightarrow bbWbj)$ could reach values of more than $1000$ fb for light $H^\pm$ and $A$. The signal $\tau\tau Wbj$ is also important in this mass region with a total cross section of almost $131$ fb. 
However, these cross sections drop rapidly once the top quark becomes off-shell ($M_{H^\pm} > m_t - m_b$). The lower panels show the results of 2HDM type-X. The signal cross sections still dominate at low Higgs masses especially the $\tau\tau Wbj$ one as expected.  
Before we investigate the signatures involving $H^\pm \rightarrow W^\pm A$ decay channel, we point out that the $h \rightarrow \gamma\gamma$ decay gains attention in the IS. This decay mode is experimentally cleanest and in some cases dominates (especially in the fermiophobic limit) before the opening of $h \rightarrow WW^*$ mode. 
In this regard, we show in Fig. \ref{gaga:signatures} $\sigma^h(pp \rightarrow \gamma\gamma WW)$ and $\sigma^h(pp \rightarrow \gamma\gamma Wbj)$ in type-I. The Former yields values up to $19$ fb while the latter reaches almost $333$ fb.
The signals $\gamma\gamma WW$ and $\gamma\gamma Wbj$ would provide a promising avenue.
Note that the total cross section for such final states are found to be negligible in type-X. 

We now turn to investigate the region below the threshold of  $H^\pm \rightarrow W^\pm A$ decay. Note that, similar to Figs. \ref{HcW:signatures} and \ref{qbHc:signatures} (upper panels), we examine the total cross sections of the similar signatures discussed above, with the only exception being the $H^\pm \rightarrow W^\pm A$ channel which involves the decays of $A$ (instead of $h$). In type-I, the $\sigma^A{(pp \rightarrow bbWW)}$ and $\sigma^A{(pp \rightarrow \tau\tau WW)}$ cross sections reach their maximum at $M_A \sim 51$ GeV. The signal cross sections $\sigma^A{(pp \to bb Wbj)}$ and $\sigma^A{(pp \to \tau\tau Wbj)}$ have sizable rates reaching almost $2180$ fb and $181$ fb at $M_A \sim 53$ GeV, respectively. In type-X, the best rates of $\tau\tau WW$ and $\tau\tau Wbj$ cross sections can yield only $0.1$ fb.
Both rates of $\gamma\gamma WW$ and $\gamma\gamma Wbj$ are found negligible in this scenario. 



In the IS, unlike the NS, both signatures arising from $H^\pm \rightarrow W^\pm h$ and $H^\pm \rightarrow W^\pm A$ decays are quite interesting. In type-I, these decay channels dominate depending on the mass of $A$, which can be considered a key parameter to focus on a specific decay channel. While in type-X, only signatures arising from $H^\pm \rightarrow W^\pm h$ can give the best reach since the $\tau\nu_\tau$ decay dominates at small values of $M_{H^\pm}$ and $M_A$. Above this region, $H^\pm \rightarrow W^\pm h$ channel dominates in the rest of the allowed mass region, almost analogous to type-I.

\section{Conclusion}
\label{sect:conclusion}
Along with confirming the SM's predictions, the finding of the Higgs particle at $125$ GeV marked the beginning of a new era of physics beyond the SM. In many of these extensions, the Higgs sector predicts additional Higgs bosons in addition to the SM-like one where the charged particle gains major attention, e.g. 2HDM. Currently, the majority of experimental searches for this charged particle concentrate on the pathways related to SM fermions. 
Due to the absence of any discernible excess, upper bounds are set on the relevant branching ratios.
For heavy $H^\pm$, the main causes of this are the huge SM backgrounds for the dominant decay mode $H^\pm \to tb$ and the relatively tiny related production cross section of $tbH^\pm$. While, for light $H^\pm$, the typical search channel $H^\pm \to \tau\nu$ has a rather low decay branching ratio. Therefore, it is in our best interest to take into account other $H^\pm$ decays in order to extend its reach at collider experiments. There have recently been initiatives to increase collider reaches by searching for charged Higgs boson in its bosonic decay modes.

In this paper, based on a 2HDM framework, we examined single charged Higgs boson production through $pp \rightarrow H^\pm W^\mp$ and $pp \rightarrow H^\pm bj$ processes at the LHC, considering the most updated experimental and theoretical constraints, assuming both normal and inverted scenarios, i.e. $h$ or $H$ being exactly the observed resonance at the LHC in 2012. We first presented the theoretical predictions for $H^\pm W^\mp$ and $H^\pm b j$ production cross sections at $\sqrt{14}~\rm{TeV}$. We demonstrated that when $\tan\beta$ is small and the condition $M_{H^\pm} < m_t - m_b$ is satisfied, the $H^\pm W^\mp$ and $H^\pm b j$ productions would have sizable cross sections. We then considered $H^\pm \to W^\pm h_i/A$ decay channels and pointed out the possible signatures that come from the aforementioned charged Higgs production and decay in both type-I and type-X of 2HDM. We specifically focused on the $b\bar{b}$, $\tau\tau$ and $\gamma\gamma$ decays for neutral Higgs states ($h_i$ and $A$) arising from the bosonic decays of charged Higgs boson. We then investigated the $bb$, $\tau\tau$ and $\gamma\gamma$ final states associated with $WW$ or $Wbj$ as possible discovery modes of a single charged Higgs boson at the LHC.

Before we close our conclusion, we point out that the $bbWW$ and $bbWbj$ signals are plagued by the huge QCD background, especially the $t\bar{t}$ one, yielding poor significance. Nevertheless, our signatures $\tau\tau WW$ and $\tau\tau Wbj$ can give the best reach since they would suppress the $t\bar{t}$ background, especially if we require at least one leptonic decay of tau leptons. 
We also suggest $\gamma\gamma WW$ and $\gamma\gamma Wbj$ as clean signatures in the inverted scenario. Such signals could provide a complementary search for a charged Higgs boson at the LHC. 

\section*{Acknowledgments}
This work is supported by the Moroccan Ministry of Higher Education and Scientific Research MESRSFC and CNRST: Projet PPR/2015/6.
MK and MO are grateful for the technical support of 
CNRST/HPC-MARWAN.

\bibliographystyle{JHEP}
\bibliography{refference}

\end{document}